\pdfoutput=1
\documentclass[epjCONF,columns]{svjour}
\usepackage[pdftex]{graphicx} 
\usepackage[varg]{txfonts} 
\usepackage[latin1]{inputenc}
\usepackage{pgf}
\def\neuone{\ensuremath{\mathchoice%
      {\displaystyle\raise.4ex\hbox{$\displaystyle\tilde\chi^0_1$}}%
         {\textstyle\raise.4ex\hbox{$\textstyle\tilde\chi^0_1$}}%
       {\scriptstyle\raise.3ex\hbox{$\scriptstyle\tilde\chi^0_1$}}%
 {\scriptscriptstyle\raise.3ex\hbox{$\scriptscriptstyle\tilde\chi^0_1$}}}}

\session-title{Presented at the 2011 Hadron Collider Physics symposium (HCP-2011),Paris, France, November 14-18 2011.}
\begin{document}
\title{Search for displaced vertexes arising from decays of new, heavy particles in 7 TeV pp collisions in ATLAS.}
\author{Frederic M. Brochu,\fnmsep\thanks{\email{frederic.michel.brochu@cern.ch}}}
\institute{H.E.P group, University of Cambridge, J.J. Thomson Avenue, CAMBRIDGE CB3 0HE, United Kingdom. \\ On behalf of the ATLAS collaboration.}
\abstract{
We present the results of a search for neutralinos decaying at a significant distance from their production point into charged hadrons and a high momentum muon, forming displaced vertexes. The analysis was performed with $33~\mathrm{pb}^{-1}$ of $pp$ collision data collected by the ATLAS experiment at the LHC in 2010 at $\sqrt{s}= 7~\mathrm{TeV}$. The poster will show some highlights of the analysis.
} 
\maketitle
\section{Introduction}
\label{intro}
\begin{figure}[h!]
\centering\includegraphics[width=5cm,height=4cm]{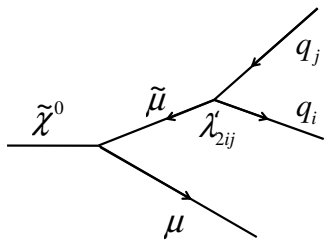}
\caption{Lightest Neutralino decay chain with a non-zero RPV coupling $\lambda'_{2ij}$.}
\label{fig:1}       
\end{figure}

In supersymmetric scenarios with R-parity violation (RPV) \cite{RefTh}, the lightest supersymmetric particle (LSP), often taken to be the neutralino $\tilde{\chi}^0_1$, is no longer stable. Decay products and LSP lifetime depend on the involved R-parity coupling type and amplitude.\\
If the coupling strength is small enough, the LSP lifetime becomes large enough to allow it to decay away from the Interaction Point (IP), leading to displaced vertex topologies.\\
We present the results of a search\cite{RefNote} for neutralinos decays into a muon and two hadronic jets via the RPV coupling $\lambda'_{2ij}$ in $33~\mathrm{pb}^{-1}$ of $pp$ collision data collected by the ATLAS experiment at the LHC in 2010 with proton beams of $3.5~\mathrm{TeV}$ each. \\
The neutralino decay chain is shown in Figure \ref{fig:1}.
\\
These neutralinos are pair-produced in the reaction \\$pp \rightarrow \tilde{q}\tilde{q} \rightarrow q \neuone q \neuone$. 
  
\section{Reconstruction of Displaced Vertexes}
\label{sec:1}
\subsection{The ATLAS detector}
\label{sec:2}
\begin{figure}
\centering\pgfimage[width=6.9cm,height=5.8cm]{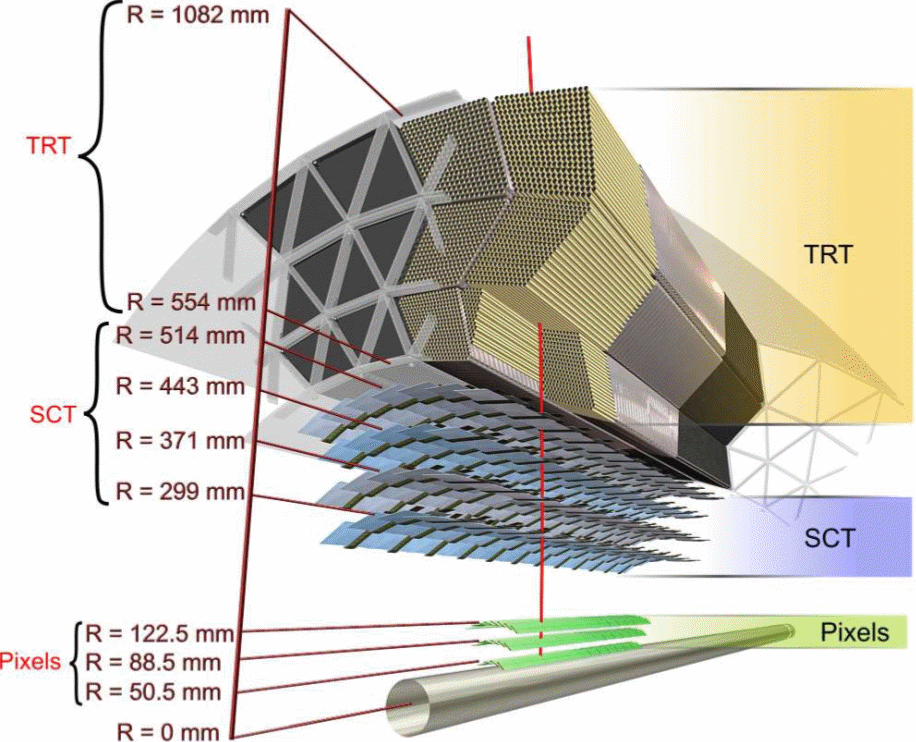}
\caption{Details of ATLAS' Inner Detector.}
\label{fig:2}       
\end{figure}

The ATLAS detector~\cite{RefAtlas} is made of a composite inner tracking system, called the Inner Detector (ID), a calorimeter system and an extensive muon spectrometer (MS). 
The ID operates in a 2~T magnetic field and provides
tracking and vertex information for charged particles in the
pseudo-rapidity range $|\eta| < 2.5$, where $\eta \equiv
-\ln\tan(\theta/2)$, and $\theta$ is the polar angle, defined with
respect to the cylindrical symmetry axis (the $z$ axis) of the detector.\\
From the IP outwards, the following components of the Inner Detector are found:
\begin{itemize}
\item the Pixel detector, made of high resolution Silicon Pixel strips forming three barrel layers and three forward disks on each side.
\item the Silicon Microstrip Tracker (SCT) made of 4 barrel layers and 9 forward disks on each side.
\item the Transition Radiation Tracker (TRT) composed of straw-tube elements interleaved with transition radiation material for electron identification.
\end{itemize}
These subdetectors are shown in  Figure \ref{fig:2}.\\
\linebreak
In this analysis, we are looking for LSP decays inside the Pixel detector, so we make an extensive use of the Inner Detector data, combined with the MS for the reconstruction of the muon.\\
\subsection{Event Selection and Reconstruction.}
\begin{figure}
\centering\includegraphics[width=7cm]{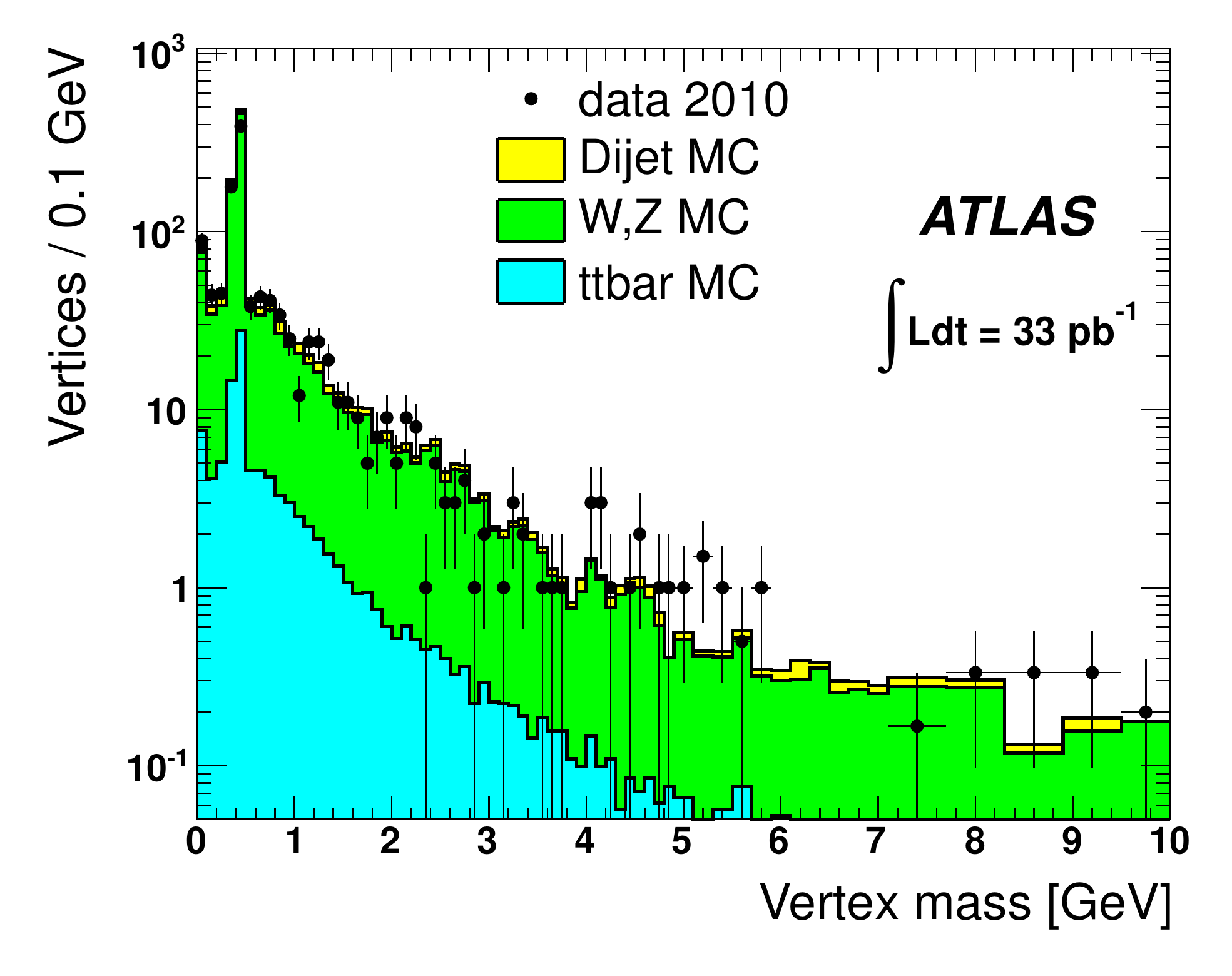}
\caption{Distribution of the reconstructed displaced vertex mass after muon and event selection cuts.}
\label{fig:3}       
\end{figure}

\begin{figure}
\centering\includegraphics[width=7cm]{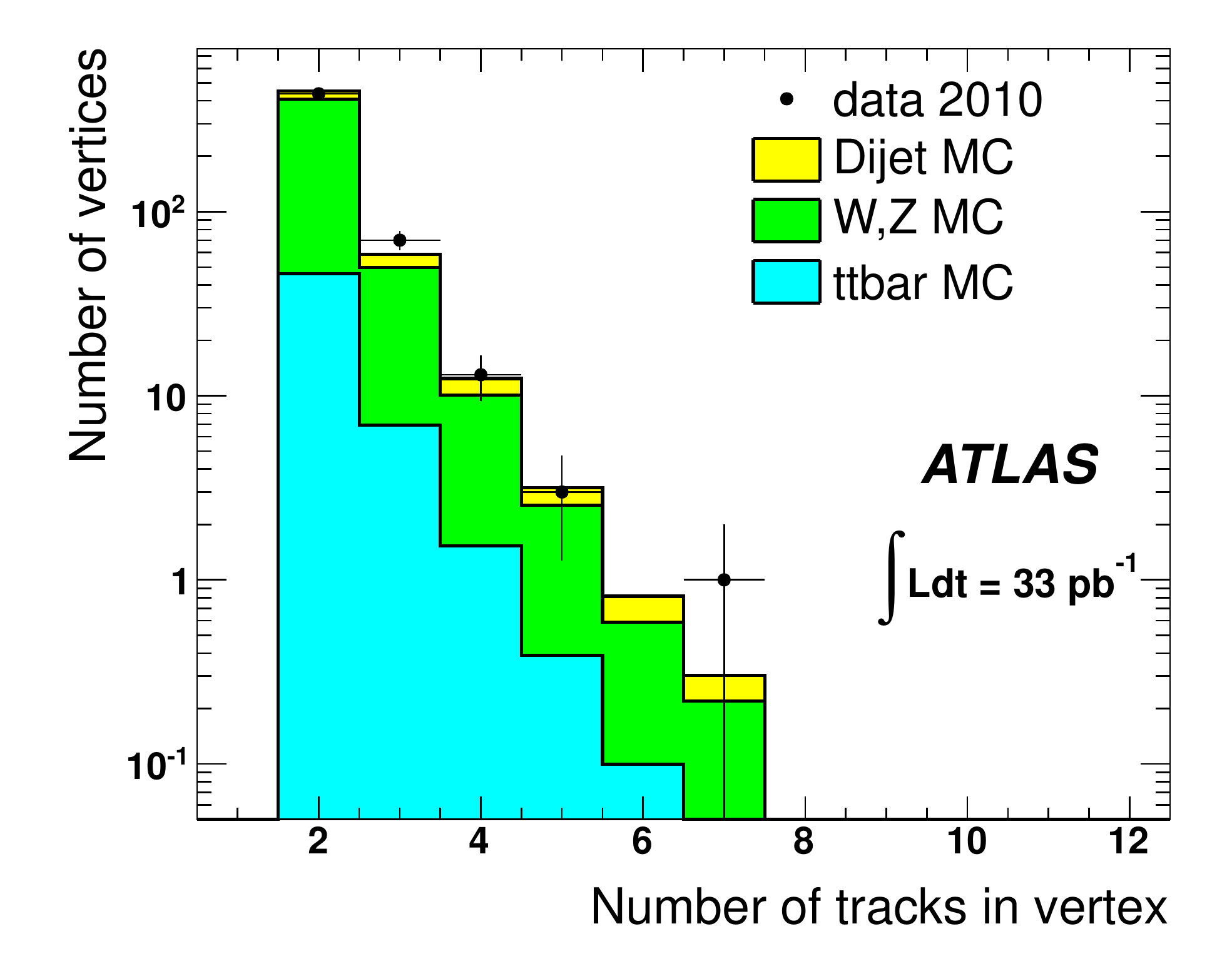}
\caption{Distribution of the number of tracks associated to the reconstructed displaced vertex after muon and event selection cuts.}
\label{fig:4}       
\end{figure}

\begin{figure}
\centering\pgfimage[width=7cm,height=5cm]{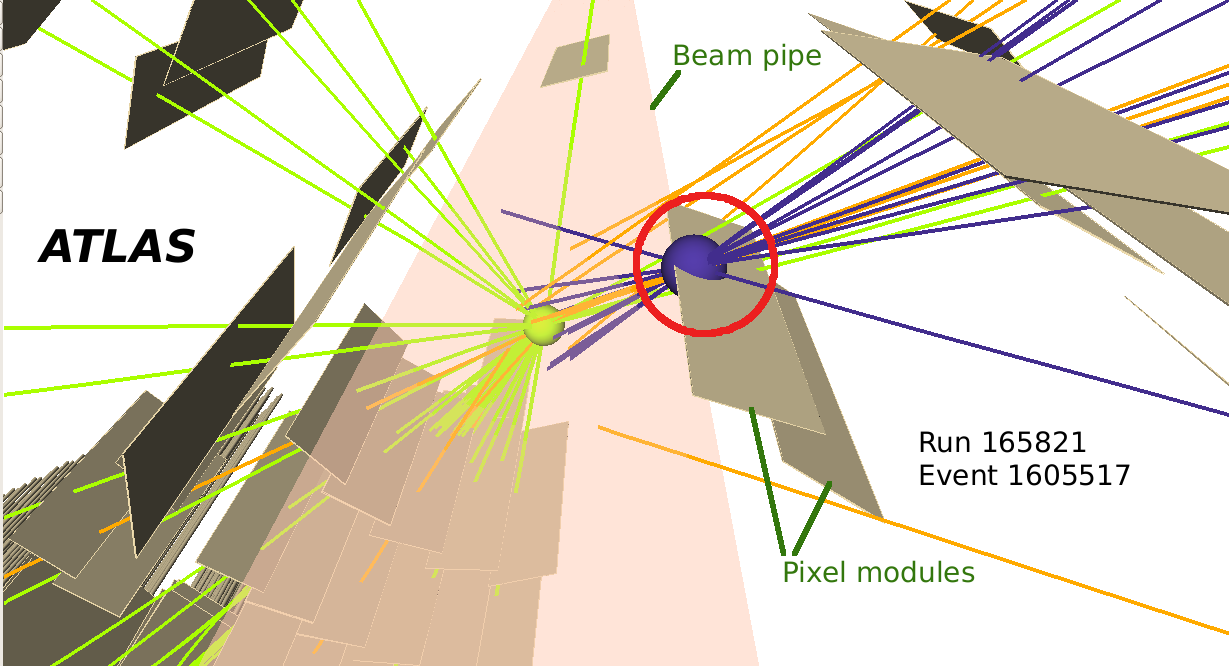}
\caption{Example of a selected displaced vertex from detector material interactions (here, in the first Pixel layer).}
\label{fig:5}       
\end{figure}

\begin{figure}
\centering\includegraphics[width=7cm]{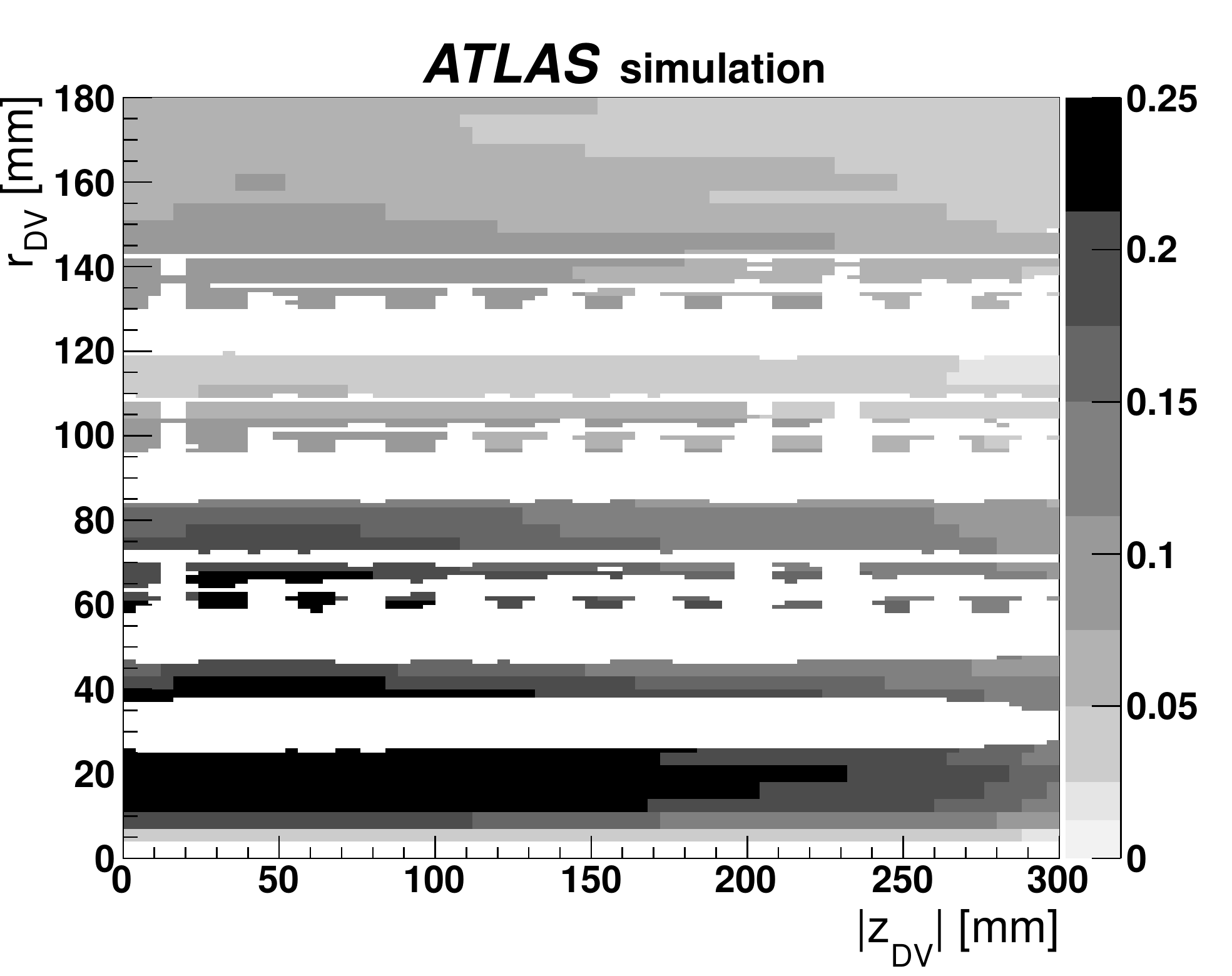}
\caption{Vertex Reconstruction efficiency as a function of the vertex position in the $R-z$ plane, after application of the detector material map veto.}
\label{fig:6}       
\end{figure}

\begin{figure}
\centering\includegraphics[width=7cm]{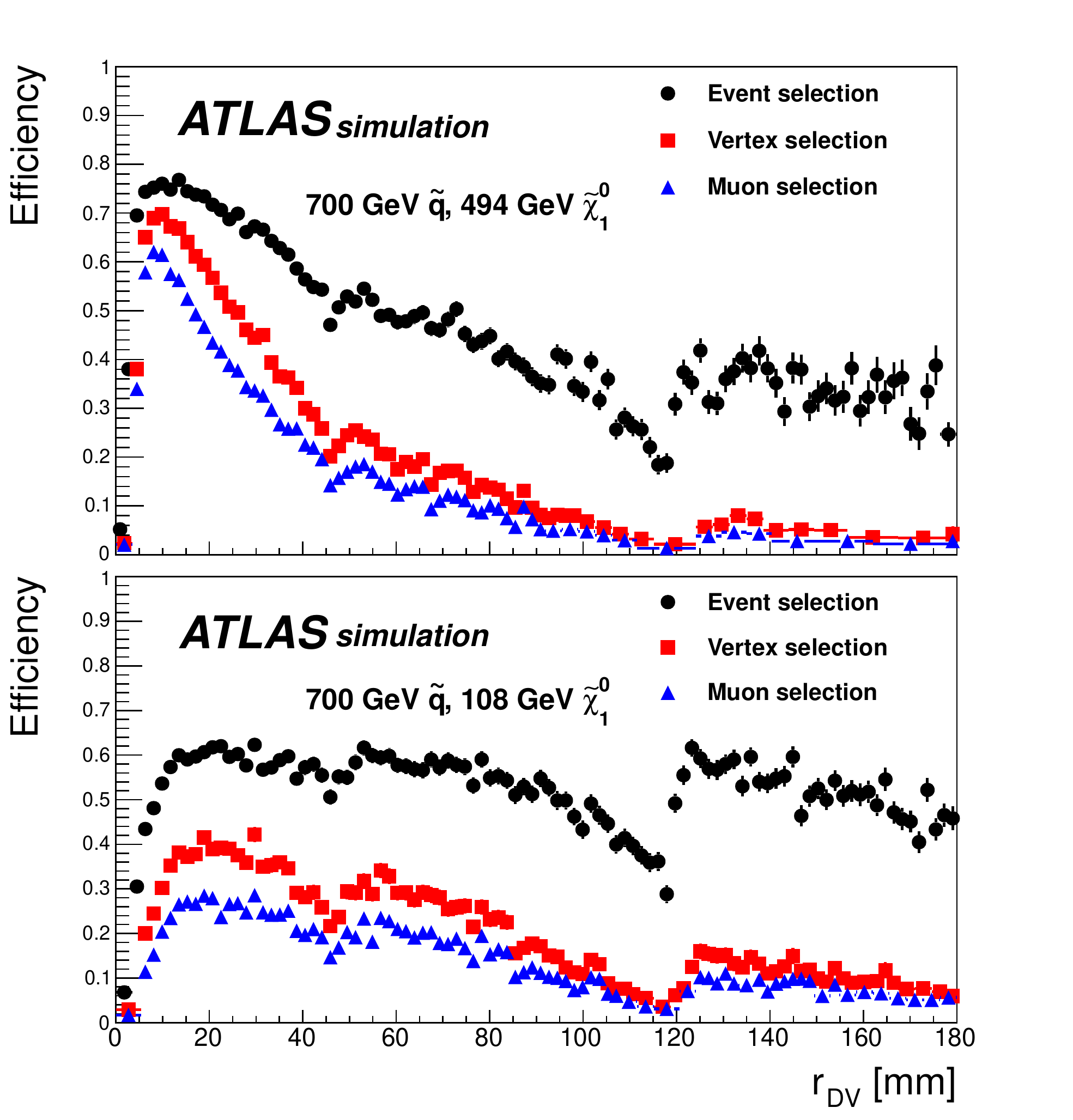}
\caption{Reconstruction efficiency as a function of the displaced vertex radius for different, cumulative set of cuts.}
\label{fig:7}       
\end{figure}

Events are required to pass the High Level muon Trigger: at least one reconstructed muon with transverse momentum $p_T> 40$ GeV. Selected events must have at least one good primary vertex, with $|z|< 200$ mm and at least 5 tracks pointing towards it.\\
Non-pointing tracks reconstructed by the Inner Detector tracking algorithms are used for the reconstruction of displaced vertexes. We use only tracks with transverse impact parameter $|d_0| > 2$ mm and $p_T > 1$ GeV.\\
These cuts form the event selection level.\\
At this point, the dominant backgrounds are events with W and Z decaying to muons. Distributions of Displaced Vertex quantities with these backgrounds overlaid are shown in Figures \ref{fig:3} and \ref{fig:4}. \\
\linebreak
A second set of cuts on vertex reconstruction observables are applied: vertex fit quality $\chi^2 < 5 * n.d.o.f$, vertex mass $M_{vtx}> 10 $ GeV and at least 4 tracks associated to the vertex. This set of cuts defines the vertex reconstruction level.\\
Some of the reconstructed vertexes at this stage are coming from interactions with the detector material, as shown in Figure \ref{fig:5}.\\
Hadronic interactions are selected and used to build detector material maps\cite{RefMap} . These maps are used to veto reconstructed vertexes, as one can see on the efficiency map reproduced in Figure \ref{fig:6}.\\
After reconstruction, the muon candidate must pass a $p_T$ cut tightened to 45 GeV and have both MS and ID data.\\ This cut represent the muon selection level.\\ 
The reconstruction efficiency as a function of the reconstructed displaced vertex radius $R_{DV}$ is shown in Figure \ref{fig:7} after applying the different cut selection levels mentioned above.
The dominant systematics errors were found to come from the following sources:
\begin{itemize}
\item muon trigger efficiency, 4.3 \%, evaluated with a $Z \to \mu \mu$ sample.
\item evolution of the muon reconstruction efficiency as a function of $d_0$, from 3.5 \% to 8 \% depending on the signal investigated, evaluated from cosmics muons.
\item vertex reconstruction efficiency, 3 to 4.3 \% evaluated from $K^0_s$ control samples.
\end{itemize}

\section{Results and Limits.}
\begin{figure}
\centering\includegraphics[width=7cm]{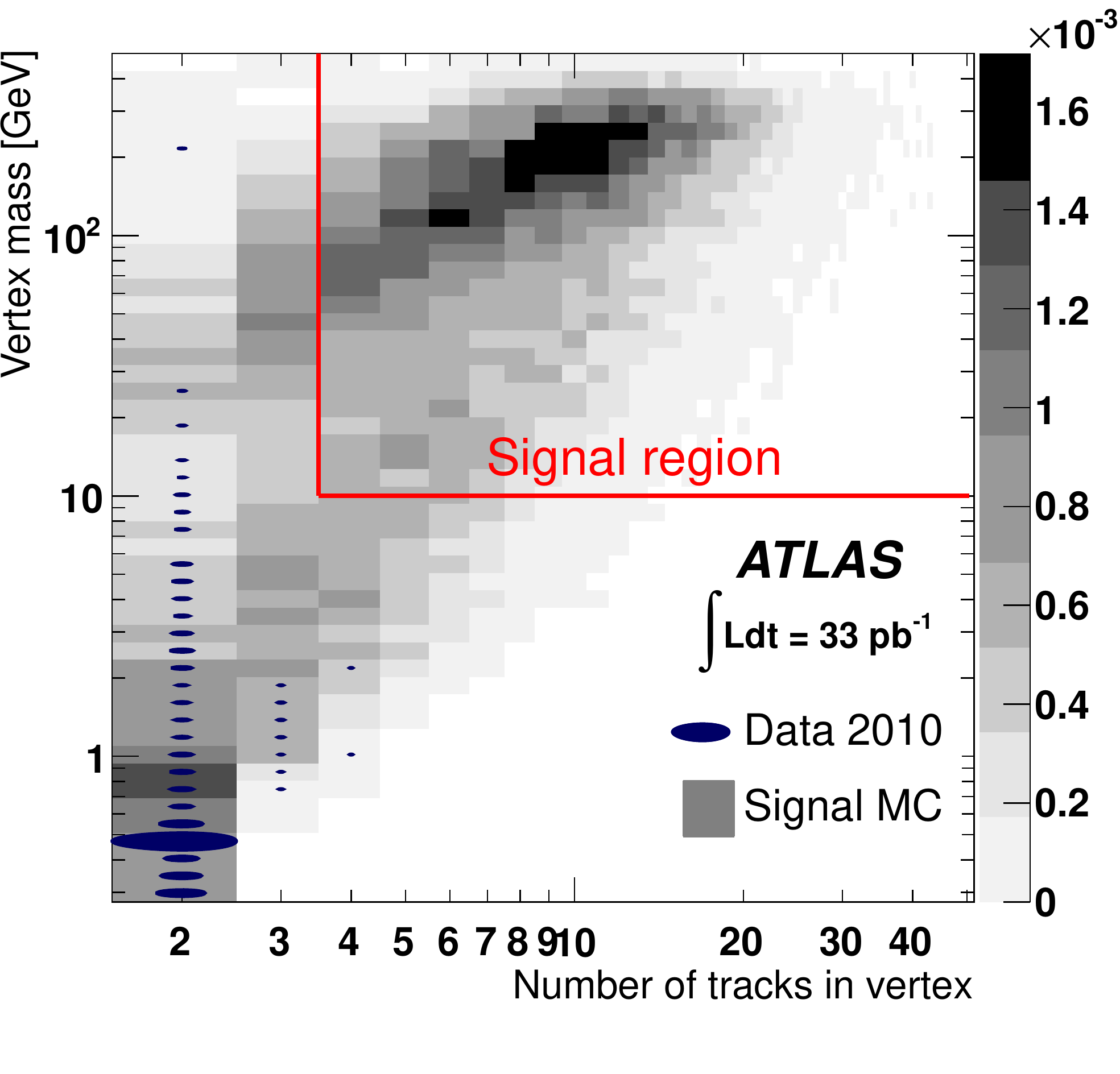}
\caption{Remaining events in data (blue, proportional ellipses) and definition of the signal region from simulation (grey level boxes). }
\label{fig:8}       
\end{figure}

\begin{figure}
\centering\includegraphics[width=7cm]{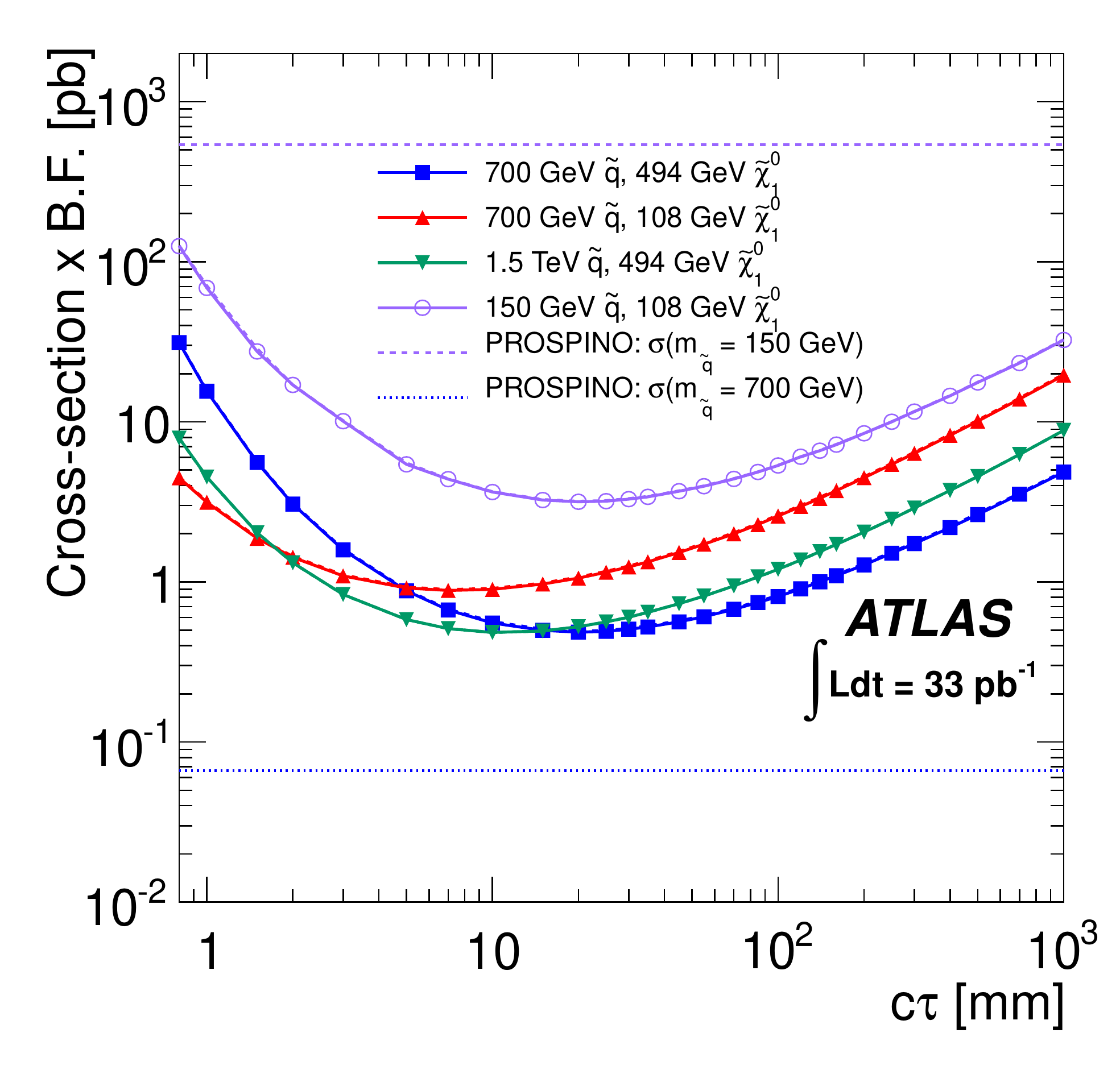}
\caption{Limits on the product $\sigma(pp \to q\neuone q\neuone) \times B.R(\neuone \to \mu q q)$ as a function of the neutralino lifetime $c\tau_{\neuone}$ for different benchmark points and event generators.}
\label{fig:9}       
\end{figure}

The vertex cuts define a signal region which is represented in  Figure \ref{fig:8}.
No data event passes the final selection, so we set limits on the product of the production cross-section and the branching ratio of the neutralino to the selected decay mode. Limits are established as a function of the lifetime of the neutralino in Figure \ref{fig:9}.

\end{document}